\documentclass[12pt,letterpaper]{article}

\usepackage{latexsym,natbib}
\usepackage{amsmath,amssymb,amsfonts,amsthm,latexsym,color,graphicx,ulem,psfrag}

\usepackage{rotating}
\usepackage[margin=52pt,small,bf]{caption}
\definecolor{myred}{rgb}{1,0.1,0.2}

\renewcommand{\mathbf}{\boldsymbol}

\renewcommand{\bf}{\bfseries}

\renewcommand{\hat}[1]{\widehat{#1}}

\newcommand{\uu}{{\mathbf{u}}}

\newcommand{\xx}{{\mathbf{x}}}
\newcommand{\yy}{{\mathbf{y}}}

\newcommand{\DD}{{\mathbf{D}}}

\newcommand{\HH}{{\mathbf{H}}}
\newcommand{\II}{{\mathbf{I}}}
\newcommand{\JJ}{{\mathbf{J}}}

\newcommand{\MM}{{\mathbf{M}}}

\newcommand{\VV}{{\mathbf{V}}}

\newcommand{\YY}{{\mathbf{Y}}}

\newcommand{\muu}{{\mbox{\boldmath $\mu$}}}

\newcommand{\SI}{{\mbox{\boldmath $\Sigma$}}}

\newcommand{\one}{{\mbox{\boldmath $1$}}}
\newcommand{\zer}{{\mbox{\boldmath $0$}}}

\begin{document}

\title{Mixed Models and Shrinkage Estimation for Balanced and Unbalanced Designs}

\author{Yihan Bao\footnote{This work began as an undergraduate
    research project}\hspace{1mm} and James G. Booth\footnote{Corresponding author: Jim.Booth@Cornell.edu}\\ Department of Statistics and Data Science, Cornell University}

\date{October 2021}

\maketitle

\begin{abstract}
The known connection between shrinkage estimation, empirical Bayes,
and mixed effects models is explored and applied to balanced and
unbalanced designs in which the responses are correlated. As an
illustration, a mixed model is proposed for predicting the outcome of
English Premier League games that takes into account both home and
away team effects. Results based on empirical best linear unbiased predictors
obtained from fitting mixed linear models are compared with fully
Bayesian predictors that utilize prior information from the previous
season. \\
Keywords: BLUP, EBLUP, empirical Bayes, football pools, MAP estimation
\end{abstract}

\baselineskip 20pt

\section{Introduction}\label{sec:intro}

The famous result due to Charles Stein \citep{stei:1955} concerning
the inadmissibility of the maximum likelihood estimator of a
multivariate normal mean with respect to squared error loss has
generated a lot of interest in shrinkage estimation in the decades
since. In particular, the connection to empirical Bayes estimation first noted by
\cite{efro:morr:1973} means that the benefits of shrinkage estimation
can be obtained using mixed effects models and, specifically, linear
mixed effects models in many situations. 

In this paper we consider estimation based on a $t\times n$ data
matrix, $\YY$, where, initially, we assume a data generating
mechanism in which the columns of $\YY$, denoted by $\yy_j$,
$j=1,\ldots,n$, are independent Gaussian vectors with mean vector
$\muu$ and covariance matrix $\SI$. In some cases, such as a balanced
incomplete block design (BIBD), some of the elements of each column of
the data matrix may be missing. In this case, and if the data is unbalanced, the value of
$\muu$ varies with $j$ (see Section~\ref{sec:epl} for a case study). 

The classical Stein estimation setting concerns the complete data case
in which $\SI=\sigma^2\II$. In this setting \cite{stei:1955} showed
that the maximum likelihood estimator (MLE) for $\muu$,
$\hat{\muu}^{MLE}=\bar{\yy}$, is inadmissible with respect to total
squared error loss for $t\geq 3$, where
$\bar{\yy}=\YY\one/n=(\bar{y}_{1\cdot},\ldots,\bar{y}_{t\cdot})^T$
denotes the mean of the rows of the data matrix. 

We review the classical setting and the connections to empirical Bayes
and linear mixed models in
Section~\ref{sec:stein}. Section~\ref{sec:rcbd} concerns the extension
of the classical results to the randomized complete blocks and
balanced incomplete block design settings in which responses within
blocks are correlated. In Section~\ref{sec:epl} we propose a linear
mixed model for predicting the outcomes (goal differences) of
individual games in the English Premier League (EPL) which takes into
account both home and away team effects. In
Section~\ref{sec:bayes} we discuss how prior information from
previous EPL seasons can be incorporated into the modeling strategy.
We conclude in Section~\ref{sec:disc} with some discussion.

\section{Classical Stein Shrinkage Estimation}
\label{sec:stein}

Suppose that $\yy_j\sim N(\muu, \sigma^2_e\II_t)$. \Citet{jame:stei:1961} showed that the estimator
\begin{eqnarray}
\label{eqn:JS}
\hat{\muu}^{JS} = \left( 1 - \frac{\sigma^2_e/n}{S_0/(t-2)} \right) \bar{\yy}\,,
\end{eqnarray}
where $S_0=||\bar{\yy}||^2$, dominates the MLE, $\bar{\yy}$, in terms of squared error loss for all values of $\muu$. The JS estimator shrinks the sample mean components towards zero (provided the multiplying factor in (\ref{eqn:JS}) is positive and less than 1). A modification, known as Lindley's estimate (or JSL) is to shrink towards the overall mean $\bar{y}_{\cdot\cdot}=\one_t^T\YY\one_n/tn$, 
\begin{eqnarray}
\label{eqn:JSL}
\hat{\muu}^{JSL} = \one_t\bar{y}_{\cdot\cdot} + \left( 1 - \frac{\sigma^2_e/n}{S/(t-3)} \right)(\bar{\yy}-\one_t\bar{y}_{\cdot\cdot})\,,
\end{eqnarray}
where $S=||\bar{\yy}-\one_t\bar{y}_{\cdot\cdot}||^2$. An unbiased estimate of
$\sigma^2_e$ is given by
$\hat\sigma^2_e=\sum_{i=1}^t\sum_{j=1}^n(y_{ij}-\bar{y}_{i\cdot})^2/t(n-1)$,
and the ratio, $\hat\sigma^2_e/\sigma^2_e$, is distributed as
$\chi^2_{t(n-1)}/t(n-1)$ independently of $\bar\yy$. This fact can be
used to show that the JSL estimator still dominates the MLE even when
$\sigma^2_e$ is estimated \cite[see e.g.,][chapter 5]{lehm:case:1998}. 

\Citet{efro:morr:1973} showed that shrinkage estimators such as (\ref{eqn:JS}) and (\ref{eqn:JSL}) can be motivated from an empirical Bayes (EB) perspective. Specifically, suppose that, conditionally on $\muu$, the columns of $\YY$ are iid $N(\muu,\sigma^2_e\II_t)$. This is the same distributional assumption made above, but now, in addition, suppose that $\muu$ is also random, and that $\muu\sim N(\zer,\sigma^2_\mu\II_t)$. Then the posterior mean of $\muu$ given the data matrix, $\YY$, is given by 
\begin{eqnarray*}
E(\muu|\YY) = \left( 1 - \frac{\sigma^2_e/n}{\sigma^2_{\mu}+\sigma^2_e/n}\right) \bar{\yy}\,.
\end{eqnarray*}
Under these modeling assumptions, an unbiased estimator of the the denominator, $\sigma^2_{\mu}+\sigma^2_e/n$, is given by $S_0/t$ resulting in an estimator the same as (\ref{eqn:JS}) except for the substitution of $t$ in place of $t-2$.

An alternative 3-parameter EB model sets $\muu = \one_t\bar\mu + \uu$, where $\bar\mu=\one_t^TE(\muu)/t$ is a fixed scalar parameter and $\uu\sim N(\zer_t,\sigma^2_u\II_t)$. In this case, the posterior mean of $\muu$ given the data matrix is given by
\begin{eqnarray}
\label{eqn:blup}
E(\muu|\YY) = \one_t\bar\mu + \left( 1 - \frac{\sigma^2_e/n}{\sigma^2_u+\sigma^2_e/n} \right)(\bar{\yy}-\one_t\bar\mu)\,.
\end{eqnarray}
Under these modeling assumptions, unbiased estimators for $\bar\mu$, $\sigma^2_e$ and  $\sigma^2_u+\sigma^2_e/n$ are given by $\bar{y}_{\cdot\cdot}$, $\hat\sigma^2_e$ and $S/(t-1)$. Substituting these values results in an estimator the same as (\ref{eqn:JSL}) except for the use of $t-1$ in place of $t-3$. An equivalent specification of this second EB model using mixed model notation is 
\begin{eqnarray}
\label{eqn:MM}
y_{ij} = \bar\mu + u_i + e_{ij}\,,
\end{eqnarray}
where $e_{ij}\sim\mbox{ iid }N(0,\sigma^2_e)$ independently of
$u_i\sim\mbox{ iid }N(0,\sigma^2_u)$. It follows that an EB version of
the JSL estimator can be obtained by fitting (\ref{eqn:MM}) using standard
mixed model software and extracting the empirical best linear unbiased
predictors (EBLUPs) for $\mu+u_i$, $i=1,\ldots,t$. In particular, the
mixed model fits in this paper were obtained using the \textit{lmer}
function from the \textit{lme4}
package in R \citep{R:2021}, with the \textit{method=REML} option. 

\section{Randomized Blocks Designs}
\label{sec:rcbd}

In this section we relax the distributional assumptions to allow for
correlation between responses. Specifically, suppose that
$\yy_j\sim\mbox{ iid }N(\muu,\sigma^2_e\II_t+\sigma^2_b\JJ_t)$, where
$\JJ_t$ denotes a $t\times t$ matrix of ones. This distributional assumption
is the same as that encountered in a classical randomized complete
blocks design (RCBD) with one complete set of treatments per block; see e.g. \citet[chapter 4]{coch:cox:1957}, whereby
\begin{eqnarray}
\label{eqn:rcbd.fixed}
y_{ij} = \bar\mu + \upsilon_i + b_j + e_{ij},
\end{eqnarray}
where $\upsilon_i$ is the deviation of the $i$th `treatment' mean from the overall mean, the $b_j$'s are random `block' effects with $b_j\sim\mbox{ iid } N(0,\sigma^2_b)$ independently of the random `errors' $e_{ij}\sim\mbox{ iid }N(0,\sigma^2_e)$. 

The normality and independence assumptions together imply that the MLE for $\muu$ is a solution of the generalized least squares (GLS) estimating equation
\begin{eqnarray}
\label{eqn:rcbd.equation}
\zer = \sum_{j=1}^n(\sigma_e^2\II_t+\sigma_b^2\JJ_t)^{-1}(\yy_j-\muu) 
\end{eqnarray}
which has solution $\hat\muu=\bar{\yy}$, that does not depend on the variance components.

Assume the goal, as before, is to estimate the treatment mean vector $\muu$ with $i$th component $\mu_i=\bar\mu+\upsilon_i$. By analogy with the EB approach described in the previous section we propose estimators based on a mixed model of the form 
\begin{eqnarray}
\label{eqn:rcbd.random}
y_{ij} = \bar\mu + u_i + b_j + e_{ij},
\end{eqnarray}
in which the `treatment' effects vector $\uu$ is distributed as $N(\zer_t,\sigma^2_u\II_t)$, and 
$\yy_j|\uu\sim\mbox{ iid }N(\one_t\bar\mu+\uu,\sigma^2_e\II_t+\sigma^2_b\JJ_t)$. The BLUP for $\one_t\bar\mu+\uu$ based on (\ref{eqn:rcbd.random}) is given by 
\begin{eqnarray}
\label{eqn:rcbd.blup}
E(\muu|\YY) = \one_t\bar\mu + \left( 1 - \frac{\sigma^2_e/n}{\sigma^2_u+\sigma^2_e/n} \right)(\bar{\yy}-\one_t\bar\mu)\,.
\end{eqnarray}
The same estimators for $\bar\mu$ and $\sigma^2_u+\sigma^2_e/n$ are valid (unbiased) here also, and an unbiased estimator of $\sigma^2_e$ is given by
\begin{eqnarray}
\label{eqn:sigma_e_hat}
\hat\sigma^2_e = \frac{1}{(t-1)(n-1)}\sum_{i=1}^t\sum_{j=1}^n(y_{ij}-\bar{y}_{i\cdot}-\bar{y}_{\cdot j} + \bar{y}_{\cdot\cdot})^2\,.
\end{eqnarray}

Responses from balanced incomplete block designs (BIBDs) are typically analyzed using a model of the form (\ref{eqn:rcbd.fixed}), the difference being that only a subset of treatments is present in each block. (An alternative view is that some block-treatment combinations are missing.) Specifically, consider designs in which $k<t$ treatments are present in each block, each treatment occurs in $r<n$ of the blocks, and each pair of treatments occur together in a block $\lambda$ times. These restrictions imply that $n=rt/k$ and $\lambda=r(k-1)/(t-1)$ \citep[chapter 9]{coch:cox:1957}. For example, if $t=10$ and $k=4$, then $n=5r/2$ and $\lambda=r/3$. The minimal design in this case has $r=6$ replicates of each treatment in $n=15$ blocks with $\lambda=2$. Let $\bar{\yy}=(\bar{y}_1,\ldots,\bar{y}_t)^T$ denote the vector of treatment means (with each component an average over $r$ replicates). Then $\bar{\yy}\sim N(\muu,a\II_t+b\JJ_t)$, where 
\begin{eqnarray*}
a = \frac{\sigma^2_e}{r} + \left( 1 - \frac{\lambda}{r} \right)\frac{\sigma^2_b}{r}
\end{eqnarray*}
and $b=(\lambda/r)\sigma_b^2/r$. Note that these formulas are also valid for the RCBD, where $r=n=\lambda$.

\noindent{\bf Theorem:} Suppose that $\bar{\yy}\sim N(\muu,a\II_t+b\JJ_t)$ for some $a>0$ and $b>0$. Let
\begin{eqnarray}
\label{eqn:muhat}
\hat{\muu} = \one_t\bar{y} + \left( 1 - \frac{a}{S/(t-3)} \right) (\bar{\yy}-\one_t\bar{y})\,,
\end{eqnarray}
where $\bar{y}=\one_t^T\bar\yy/t$. Then
\begin{eqnarray}
\label{eqn:EMSE}
E\left\{ ||\hat\muu-\muu||^2 \right\} \leq E\left\{ ||\bar{\yy}-\muu||^2 \right\}
\end{eqnarray} 
for all $\muu$.

The proof is given in the Appendix. 

The theorem implies that the estimator (\ref{eqn:muhat}) dominates the
treatment mean vector, $\bar{\yy}$, in both the RCBD and
BIBD. However, we note here that, while $\bar{\yy}$ is the MLE for
$\muu$ in a RCBD, this is not the case for a BIBD. To see this, let
$\MM_j$ denote the $k\times t$ matrix that contains the rows of a
$t\times t$ identity matrix corresponding to the $k$ treatments that
are present in the $j$th block. Then the GLS estimating equation for
$\muu$ for this linear model corresponding to (\ref{eqn:rcbd.equation}) is 
\begin{eqnarray}
\label{eqn:bibd.equation}
\zer &=& \sum_{j=1}^n\MM_j^T\left[ \MM_j(\sigma_e^2\II_t+\sigma_b^2\JJ_t)\MM_j^T \right]^{-1}(\yy_j-\MM_j\muu)\,.
\end{eqnarray}
which has solution
\begin{eqnarray}
\label{eqn:bibd.solution}
\hat\muu^{MLE} = \left( \sum_{j=1}^n\MM_j^T\left[ \MM_j(\sigma_e^2\II_t+\sigma_b^2\JJ_t)\MM_j^T \right]^{-1}\MM_j \right)^{-1}
\sum_{j=1}^n\MM_j^T\left[ \MM_j(\sigma_e^2\II_t+\sigma_b^2\JJ_t)\MM_j^T \right]^{-1}\yy_j\,,
\end{eqnarray}
that reduces to $\bar\yy$ in the RCBD case (where $\MM_j=\II_t$), but
not in the BIBD case. Intuitively, the block effects cancel out in the
treatment means in RCBD but not in the BIBD because, in the latter
case, different treatments occur in different subsets of blocks. See
\citet[chapter 8]{mose:1996} for a similar development of the BIBD
analysis. 

\begin{figure}[!ht]
\begin{center}
\includegraphics[height=9cm,width=16cm]{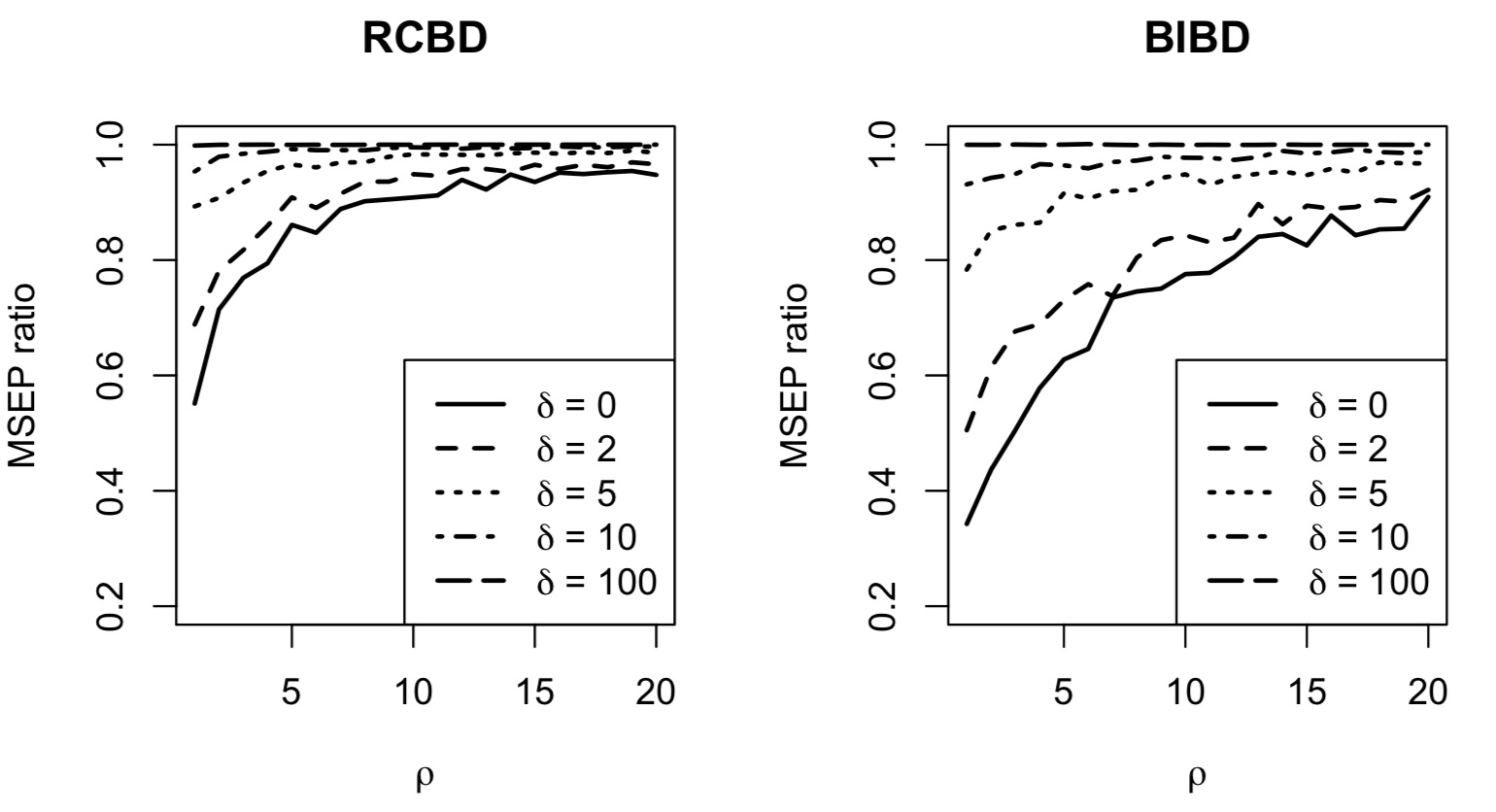}
\caption{Empirical BLUP versus MLE mean squared error of prediction ratios for varying values of $\rho=\sigma_b^2/\sigma_e^2$, and ranges for treatment means.}
\label{fig:sim}
\end{center}
\end{figure}

Figure~\ref{fig:sim} displays the results of a simulation study. The
left panel concerns a RCBD setup with $t=21$ treatments arranged in
$n=10$ blocks. Data was simulated from the model
(\ref{eqn:rcbd.fixed}) with $\sigma_e^2=10$, the ratio of block to error variance,
$\rho$, varying from 1 to 20, and the treatment means are equally spaced
over a range $-\delta/2$ to $\delta/2$, for values of $\delta$ equal
to 0, 2, 5, 10 and 100. The right panel concerns a BIBD setup with 
$t=21$, $k=7$, $n=30$ and $\lambda=3$. Thus, the total sample size in each
data set is 210, the same for both the RCBD and BIBD simulations. The
y-axis in each plot is the ratio of mean squared error of prediction
(MSEP) for empirical BLUPs and the MSEP for the MLE, based on the
model (\ref{eqn:rcbd.random}), and averaged over 100 simulated
datasets. The EBLUPs were computed using the \textit{lmer} function
from the \textit{lme4} package in R \citep{R:2021} by fitting a mixed
model with both blocks and treatments as random factors, extracting
the random treatment effects and adding the intercept.


The figure reveals that
the EBLUP dominates the MLE with a smaller mean squared
error in both RCBD and BIBD cases, but this advantage
weakens as the ratio of block to error variance, $\rho$,
increases. Also, with the same $\rho$ and treatment mean vector,
$\delta$, the relative efficiency of the EBLUP compared to the
MLE is higher for the BIBD than for RCBD. 

\section{Application}
\label{sec:epl}

The English Premier League consists of 20 teams, three of
which are relegated each year and replaced the next season by three teams
from the English League Championship. During the course of a season
each team plays every other team twice, once at home and once away. Here we
attempt to use a mixed effects model and shrinkage estimation to
predict the outcomes of remaining games based on each team's performance in their
first 7 home and away games. Since each team plays 19 home games
the initial 7 home games for each team represent (approximately) 14
games per team in total,
just over one third of a season.\footnote{The 7th home game for team
  A might be the 8th away game for team B, in which case the model fit
  for team B is based on 7 home and 8 away games, 15 in total.} We use a model with the
response being the difference, $y$, between the number of goals scored
by the home team and the number scored by the away team because the
difference between the number of goals scored by two teams is
approximately normal. This follows from the fact,
illustrated later in the Discussion, that the  
number of goals scored per game by each team is well approximated by a
Poisson distribution, and the difference between two independent
Poisson variables with means typical for the EPL is approximately normal. 

The model for a particular game
between (home) team $i$ and (away) team $j$ is
\begin{eqnarray}
  \label{eqn:ha.model}
  y_{ij} = \mu + h_i - a_j + e_{ij}\,,
\end{eqnarray}
where $h_i$ is a random effect representing the strength of team $i$
at home and $a_j$ is a random effect representing of the strength of team
$j$ away. 
We assume that $h_i\sim \mbox{iid } N(0,\sigma^2_h)$, $a_j\sim
\mbox{iid } N(0,\sigma^2_a)$, and $e_{ij} \sim \mbox{iid }
N(0,\sigma^2_e)$ for $i$ and $j$ in $\{1,2,\ldots,20\}$,
but excluding cases in which $i=j$. The data used to train the model 
is unbalanced because different teams play different opponents in
their first 7 home games. 

\begin{figure}[htbp]
\begin{center}
\includegraphics[height=9cm,width=14.4cm]{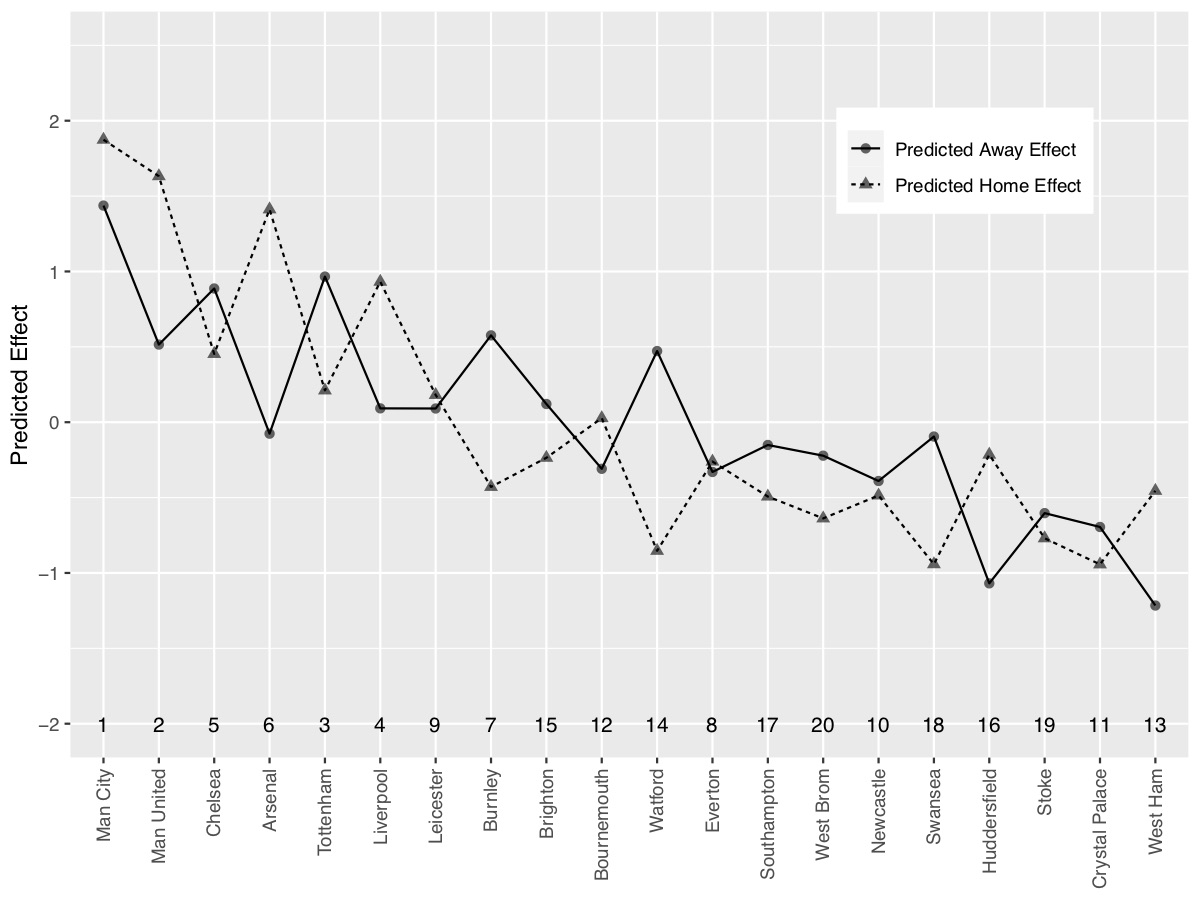}
\caption{Empirical BLUPS for each team's home and away effects for the 2017-18 season, and their actual end of season ranking.}
\label{fig:ha}


\includegraphics[height=9cm,width=14.4cm]{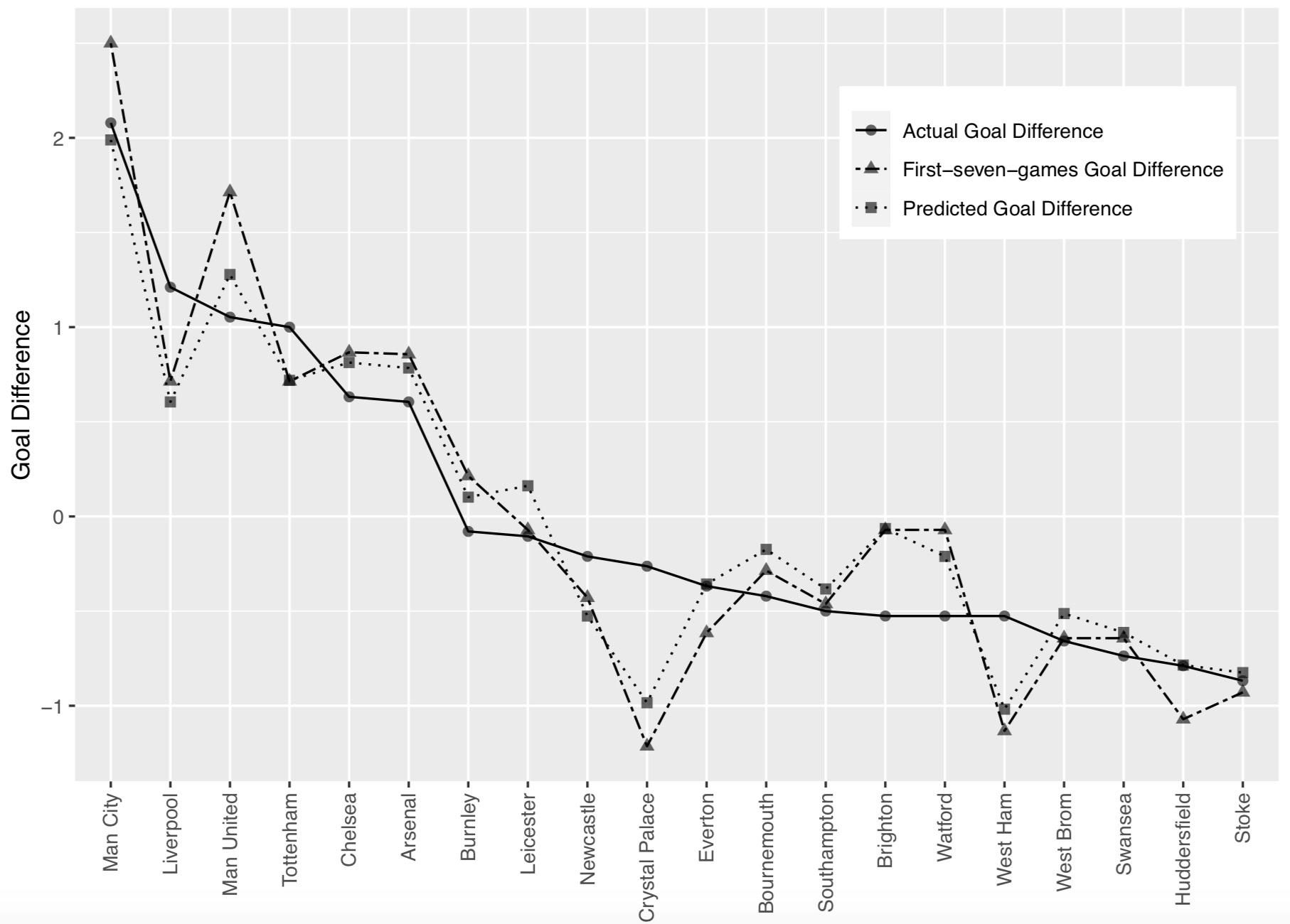}
\caption{Actual and predicted end-of-season goal difference, and actual first-seven-games goal difference for each team for the 2017-18 season.}
\label{fig:2017}
\end{center}
\end{figure}

Figure~\ref{fig:ha} shows the EBLUPs for the home and away effects for
the 2017-18 season, where the teams are ordered by the sum of their
predicted home and away effects. Thus, for example, Manchester
United's predicted home effect is 1.632, whereas West Ham United's away
effect is -1.216. If the end of season rankings of the 20 teams are
predicted using this ordering, then the mean
absolute error in predicted rankings is 3 for the 2017-18 season, and
the correlation with the actual end of season ranking is 0.788. 
Figure~\ref{fig:2017} shows the actual and predicted
end-of-season goal difference for each team in 2017-18. The predictions
are computed by averaging over the 24 games that were not used to fit
the home-and-away model (\ref{eqn:ha.model}) and combining this
average with the known results for the 14 games that were used to fit
the model. The actual goal differences based on the games used to fit
the model and the end of season goal differences for each team are also plotted for
comparison. The teams are sorted by the end of season goal
differences. 


In general we predict a win for the home team if the difference between the
home and away effects are greater than a threshold, $d$, a loss if the
difference is less than $-d$, and a tie if neither of these
inequalities hold. Predicting ties is notoriously difficult in
football. There is a long history of betting on the outcome of
football games in Britain dating back to the 1920's. The most famous
`football pool' involved picking a line of eight from the
upcoming weekend's games from all British professional
leagues. One point was awarded for picking a home win, 1.5 points for
an away win, 2 points for a 0-0 tie, and 3 points for a scoring
tie. Millions of people took part in the pool each week resulting in a
very large payout for the winners with the highest points total. The
winners often had little knowledge of football and essentially picked
their line at random. We used model (\ref{eqn:ha.model}) to pick ties,
with $d=0.25$ for all the games (after the first 7 home games)
in the 17 EPL seasons from 2001-2 to 2017-18, resulting
in overall success rate of 0.290. The value $d=0.25$ was chosen
based on a grid search, because this choice led to 1135 predicted
draws which was the closest to the actual number, 1022.

The proportion of games that were
ties during this period was 0.250, one-third of which were 0-0
ties. The complete results are given in Table~\ref{tab:results}. Based
on this historical data the estimated win, loss and draw probabilities
for the home team are 0.473, 0.277 and 0.250 respectively. Thus, if the
result of a game is predicted at random, with 1/3 probability for each
outcome, and factoring in the fact that one-third of draws are 0-0, the
expected number of points using the scoring system described in the
previous paragraphs is
\[ (1)(0.473)/3+(1.5)(0.277)/3+(2)(0.25)/9+(3)(0.25)(2/9)=0.518\,. \]
Thus, the average score for a better who chooses a line of eight games
at random is 4.144.
In contrast, the model (\ref{eqn:ha.model}) correctly predicts wins, losses and draws with
relative frequencies 0.322, 0.076 and 0.081 respectively. Also, among
the correctly predicted draws one third were 0-0\footnote{Of the 329
  draws that the model correctly predicted, 110 were no-score
  draws. Among the 1022 actual draws, 323 were no-score draws.}. Thus, the expected
number of points based on predicting outcomes using the model
(\ref{eqn:ha.model}) is
$(1)(0.322)+(1.5)(0.076)+(2)(0.081)(1/3)+(3)(0.081)(2/3)=0.652$, or
5.216 for an eight game line. Although this is an improvement over
random guessing, a score near
the mean would never win the pool. Rather it is scores in the right tail of the
distribution that determine the winner. By simulating lines of eight games
using the multinomial distributions for the four outcomes (win, loss,
no-score, and score-draw) one can establish that the model
produces a score distribution that is stochastically larger than
random guessing. So, using shrinkage estimation via a normal theory
mixed model does improve the odds of winning, but we acknowledge that
even doubling a random win probability on
the order of $10^{-6}$, or even $10^{-7}$, would not be very helpful to
a typical bettor. 

\begin{table}
\caption{Actual versus predicted results for seasons 2001-2 to 2017-18
based on model (\ref{eqn:ha.model}). (D=draw, L=loss, W=win for the
home team)}
\label{tab:results}
\centering
\begin{tabular}[t]{l|c|c|c||c}
\hline
& Predicted.D & Predicted.L & Predicted.W & Actual Totals\\
  \hline
  \hline
Actual.D & 329 & 162 & 531 & 1022\\
\hline
Actual.L & 380 & 311 & 439 & 1130\\
\hline
Actual.W & 426 & 187 & 1315 & 1928 \\
\hline
  Predicted Totals & 1135 & 660 & 2285 & 4080 \\
  \hline
\hline
\end{tabular}
\end{table}

\section{Use of Prior Information}
\label{sec:bayes}

Since there is data on EPL results from multiple seasons, it is
natural to think about incorporating prior information about model
parameters (specifically, the variance components)
based on data from previous seasons. Consider the model
(\ref{eqn:ha.model}) applied to a complete EPL season. If we ignore the
fact teams do not play themselves, the design consists of two crossed
(random) home and away team factors. The expected mean
square (EMS) for the home team factor, after adjusting for the away team,
is therefore approximately $19\sigma^2_h+\sigma^2_e$. Here we
use a multiplier of 19 rather than 20 to account, at least
approximately, for the fact that
teams do not play themselves. Similarly, the
EMS for the away team factor is approximately
$19\sigma^2_a+\sigma^2_e$. It follows that the EMS estimates of the
variance components have approximate marginal distributions given by 
$\hat\sigma^2_e\sim \sigma^2_e\chi^2_{d_e}/d_e$, $\hat\sigma^2_h\sim
\sigma^2_h\chi^2_{d_h}/d_h$, and $\hat\sigma^2_a\sim \sigma^2_a\chi^2_{d_a}/d_a$,
where $d_e=20\times 19-19-19-1=341$, and $d_h$ and $d_a$ are obtained using
Satterthwaite's approximation \citep{satt:1946}. 


\begin{figure}[!ht]
\begin{center}
\includegraphics[height=9cm,width=14.4cm]{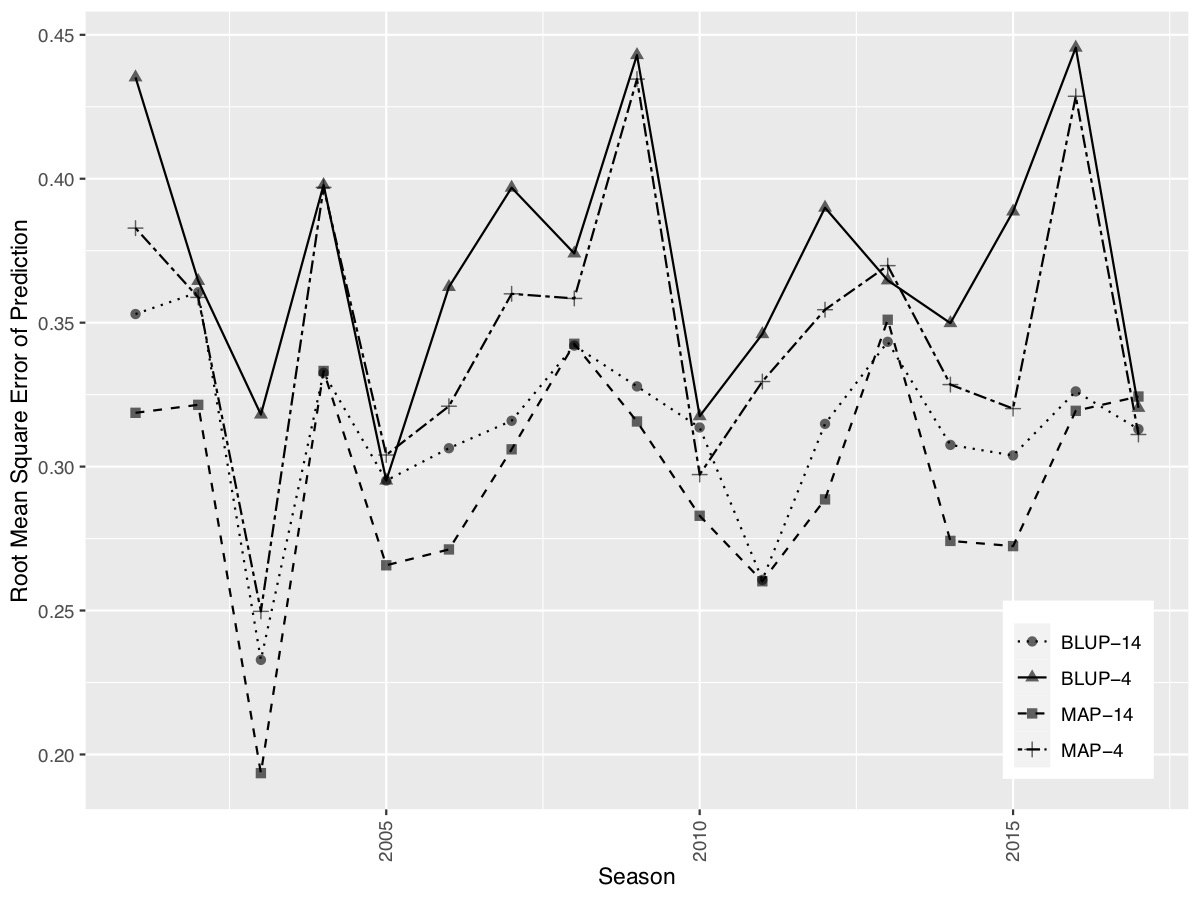}
\caption{Comparison of RMSEP for BLUP and MAP predictors based on models (\ref{eqn:ha.model}) and (\ref{eqn:MM}) for seasons beginning in 2001 to 2017.}
\label{fig:Bayes}
\end{center}
\end{figure}

If $X\sim\sigma^2\chi^2/d$ and we assign an improper
prior to $\sigma^2$ proportional to $1/\sigma^2$, then the posterior for $\sigma^2$
is inverse-Gamma with shape $\alpha=d/2$ and scale
$\beta=dX/2$.
It follows that an approximate maximum a posteriori (MAP) estimate of
$\sigma^2$ under these modeling assumptions is the posterior mode,
given by $\beta/(\alpha+1)=\frac{d/2}{d/2+1}X$.
Figure~\ref{fig:Bayes} shows the RMSEP values for EBLUP and MAP
predictors of goal differences for every game in each season after the
models were trained on the first 7 home games for each team.
The MAP predictors were computed using
the {\it blme} package in R \citep{R:2021}. There is noticable
improvement using both home and away team factors (model~\ref{eqn:ha.model})
relative to only the home team factor (model~\ref{eqn:MM}). Also the
inclusion of prior information about the variance components based on
data from the previous season improves the predication accuracy in
most years. This is reflected in the ability to predict the end of
season rankings with the mean absolute deviations between actual and
predicted rankings equal to 2.541, 2.735 for MAP and EBLUP
respectively based on
model (\ref{eqn:ha.model}), and 3.135 and 3.524 for MAP and EBLUP
based on model (\ref{eqn:MM}), over the course of 17 seasons. The
incorporation of prior information from the previous season also
improve the prediction accuracy for ties to 0.298.

\section{Discussion}
\label{sec:disc}
\begin{figure}[!ht]
\begin{center}
\includegraphics[height=9cm,width=14.4cm]{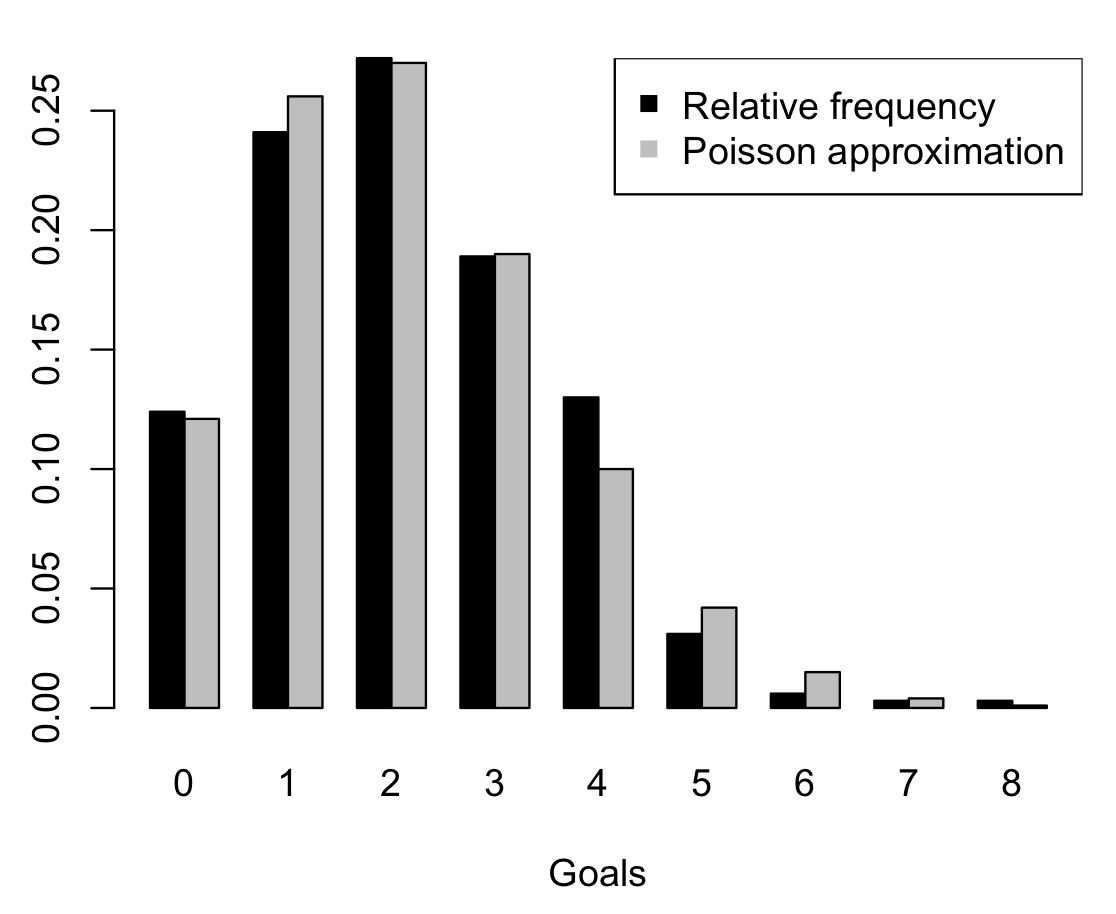}
\caption{Distribution of Goals scored by Manchester United in home EPL games between 2001 and 2018 and Poisson approximation.}
\label{fig:manu}
\end{center}
\end{figure}

\begin{figure}[!ht]
\begin{center}
\includegraphics[height=9cm,width=16cm]{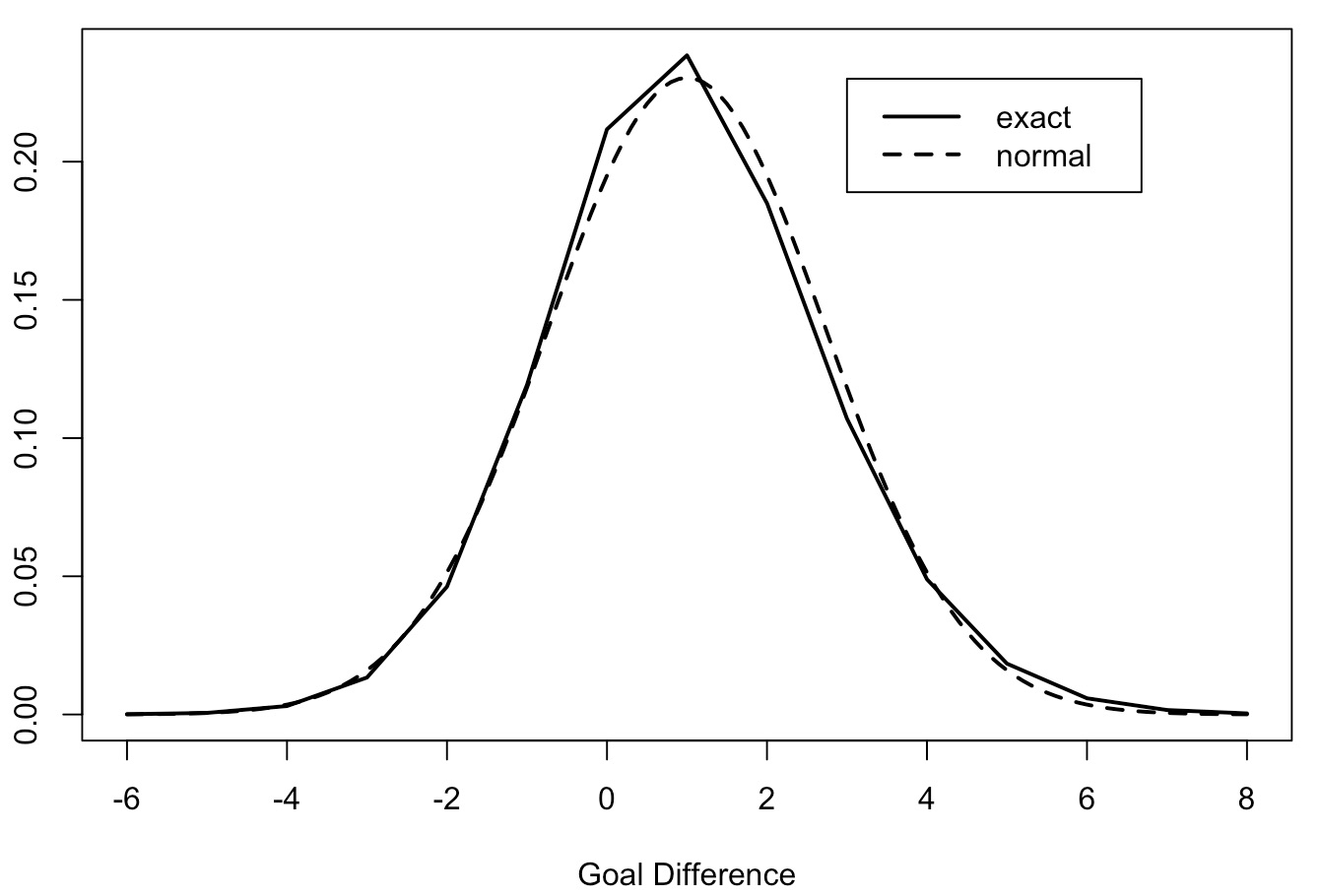}
\caption{Normal approximation to the exact distribution of the
  difference between two independent Poisson variables with means 2
  and 1 respectively.}
\label{fig:normal}
\end{center}
\end{figure}

Perhaps the most famous example of Stein shrinkage estimation concerns the
prediction of baseball batting averages for 18 major
league players based on their first 45 at bats during the 1970 season
\citep{efro:morr:1975,efro:morr:1977}. In that setting the data
consist of binomial counts with the individual hit probabilities being the
quantities of interest. Since the variance of a binomial count depends
on the success probability, the constant variance assumption
in the classical James-Stein and James-Stein-Lindley estimators given
in (\ref{eqn:JS}) and (\ref{eqn:JSL}) is
violated. \citeauthor*{efro:morr:1975} got around this issue using the
variance stabilizing, arcsine square root transformation of the
binomial proportions which results in transformed responses
that are approximately normal with variances equal to
1. In the case of goal differences in football matches it is not clear
that a transformation will help. For example, one might model the
number of goals scored by the home and away teams separately using
independent Poisson distributions, which would suggest the use of a
square root transformation. 
In fact, the Poisson distribution does appear to be a good model for
the number of goals scored by a particular team. For example,
Figure~\ref{fig:manu} shows the relative frequency distribution of the
number of goals scored in EPL home games by Manchester United from
2001 to 2018, and the corresponding probabilities for a Poisson
distribution with the same mean, 2.14. Similar excellent fits of the
Poisson distribution is found for most teams. However, the average number of goals
scored per game is about 1 for bad teams and 2 for good teams. Over
this range the square root transformation does not yield a good normal
approximation, although it is close to being variance stabilizing. On
the other hand, the difference of two independent Poisson variables
with means in this range is well approximated by a normal.
This fact is illustrated in Figure~\ref{fig:normal}
where the exact distribution of the difference between two independent
Poisson variables with means 2 and 1 is compared to a normal
distribution with the same mean and variance. The
normal approximation is even better for more evenly matched teams. 
Since different teams have different means the constant variance
assumption underlying the JSL estimator is violated. However,
with the means restricted to the range 1 to 2 the standard deviation
of the difference only varies from $\sqrt{2}$ to 2.
Of course the independence assumption is also questionable, as is
the assumption that the team effects remain constant over the
season. Thus, there is scope for improvements to the modeling approach
described in this paper. 


\section{Appendix}

\noindent{\bf Proof of Theorem}: Suppose $\yy\sim N(\muu_y,
a\II_t+b\JJ_t)$, where $a>0$ and $b>0$. Let $\VV=a\II_t+b\JJ_t$ and
note that $\VV=\HH\DD\HH^T$ where $\HH$ is the $t\times t$ Helmert
matrix and $\DD=\mbox{diag}(a+tb,a\one_{t-1}^T)$. Define
$\xx=\VV^{-1/2}\yy$, so that $\xx\sim N(\muu_x=\VV^{-1/2}\muu_y,\II_t)$. 
It follows that the JSL estimator for $\muu_x$, given
by
\begin{eqnarray*}
  \hat\muu_x^{JSL} = \one_t\bar{x} + \left( 1 - \frac{1}{S_x/(t-3)} \right)(\xx-\one_t\bar{x})\,,
\end{eqnarray*}
where $\bar{x}=\one_t^T\xx/t$ and $S_x=||\xx-\one_t\bar{x}||^2$, has the property that 
\begin{eqnarray}
  \label{eqn:inequal}
  E_{\muu_x}(||\hat\muu_x-\muu_x||^2) \leq E_{\muu_x}(||\xx-\muu_x||^2)
\end{eqnarray}
for all $\muu_x$. 

\medskip\noindent Now, the spectral decomposition of $\VV$ implies that
$\VV^{1/2}\one_t\bar{x}=\one_t\bar{y}$. Furthermore, 
\begin{eqnarray}
  \label{eqn:inverse}
  \VV^{-1}=\frac{1}{a}\left( \II_t-\frac{b}{a+tb}\JJ_t \right)
\end{eqnarray}
and $\JJ_t(\yy-\one_t\bar{y})=\zer$, together imply that $S_x = S_y/a$
and hence that 
\begin{eqnarray*}
  V^{1/2}\hat\muu_x^{JSL} = \one_t\bar{y} + \left( 1 -
  \frac{a}{S_y/(t-3)} \right) (\yy-\one_t\bar{y}) = \hat\muu_y\,.
\end{eqnarray*}
It follows from (\ref{eqn:inequal}) that 
\begin{eqnarray*}
  E_{\muu_y}\left\{(\hat\muu_y-\muu_y)^T\VV^{-1}(\hat\muu_y-\muu_y)\right\}
  \leq E_{\muu_y}\left\{(\yy-\muu_y)^T\VV^{-1}(\yy-\muu_y)\right\}
\end{eqnarray*}

\medskip\noindent The result now follows by noting that (\ref{eqn:inverse}) implies
\begin{eqnarray*}
  a(\yy-\muu_y)^T\VV^{-1}(\yy-\muu_y) = ||\yy-\muu_y||^2 +
  \frac{tb}{a+tb}(\bar{y}-\bar\mu_y)^2\,,
\end{eqnarray*}
where $\bar\mu_y=\one_t^T\muu_y/t$, and
\begin{eqnarray*}
  a(\hat\muu_y-\muu_y)^T\VV^{-1}(\hat\muu_y-\muu_y) =
  ||\hat\muu_y-\muu_y||^2 + \frac{tb}{a+tb}(\bar{\hat\mu}_y-\bar\mu_y)^2\,,
\end{eqnarray*}
where $\bar{\hat\mu}_y=\one_t^T\hat\muu_y/t=\bar{y}$. 

\bibliographystyle{natbib}
\bibliography{refs}

\end{document}